\documentclass[twocolumn]{jpsj3}
\usepackage{txfonts}
\usepackage{color}
\usepackage{graphicx}
\usepackage{amsmath}
\usepackage{amsfonts}
\title{An extension of estimation of critical points in ground state for random spin systems}

\author{Masayuki Ohzeki\thanks{mohzeki@tohoku.ac.jp}, Yuta Kudo, Kazuyuki Tanaka}
\inst{Department of Graduate School of Information Sciences, Tohoku University, Sendai, Miyagi 980-8579, Japan} 

\abst{
Most of the analytical studies on spin glasses are performed by using mean-field theory and renormalization group analysis.
Analytical studies on finite-dimensional spin glasses are very challenging.
In this short note, a possible extension of the approaches on the phase transition in spin glasses is demonstrated.
To validate our extension, we compared our estimates on the critical points with the existing numerical results.
}

\begin{document}

\maketitle

The estimation of the location of the critical points in finite-dimensional spin glasses is essential information for the determination of the critical behavior in the phase transition and to estimate the critical exponents.
Over many years of analyses of the finite-dimensional spin glasses, several controversial discussions on critical exponents have appeared repeatedly \cite{Binder1986}.
A very rare establishment on the estimation of the location of the critical points is by the duality analysis \cite{Nishimori2002,Mailard2003,Nishimori2006}.
To include the disorder effect of the spin glasses, the replica method and partial trace of spin variables give higher precision of the straightforward application of the duality analysis \cite{Ohzeki2008,Ohzeki2009,Ohzeki2014,Nishimori2010}.
The method can be extended to the case with loss of the interactions (bond dilution) \cite{Ohzeki2012,Ohzeki2012a,Ohzeki2013} and that beyond the Ising model with self-duality in the corresponding model without disorder \cite{Ohzeki2009a,Hector2012,Ohzeki2014}.
The absence of the spin glass transition on the self-dual lattices, including the standard square lattice, can be shown by the duality analysis \cite{Ohzeki2009sg}.
These methods deal with the case of the finite-temperature region, especially on the Nishimori line \cite{Nishimori1981} and the related region by using the special gauge symmetry \cite{Georges1987}.
In addition, the phase boundary close to the Onsager point can be also obtained by the duality analysis \cite{Ohzeki2011,Ohzeki2014}.
Although a challenging analysis \cite{Ohzeki2013proc} was performed on the ground state, it ended up failure.

One of the successful approaches different from the duality analysis for estimating the critical points in the finite-dimensional spin glasses is the frustration analysis proposed by Miyazaki \cite{Miyazaki2013}.
This method is applicable to the estimation of the critical point in the ground state and demonstrates a good comparison with the existing numerical results.
As pointed out by Nishimori \cite{Nishimori1986}, the geometry of the frustration should be a key point in understanding the nature of the phase transition in the ground state.
Miyazaki conjectured that the increase in frustration by induced disorder interactions has some information on the phase transition of the finite-dimensional spin glasses in the ground state.

In the present note, we report an extension of the estimation of the critical points in the ground state to the case with bond dilution.
To check our extension, we compared the estimation with the existing results \cite{Stace2010}.

We deal with the following standard model of the finite-dimensional spin glass, the so-called Edwards-Anderson model \cite{Edwards1975}:
\begin{equation}
H = - \sum_{\langle ij \rangle }J_{ij} S_i S_j,
\end{equation}
where $S_i = \pm 1$ and $\langle ij \rangle$ denotes the nearest neighboring pairs on the lattice.
The strength of the interaction $J_{ij}$ follows the distribution function 
\begin{equation}
P(J_{ij}) = (1-q)\left( (1-p) \delta(J_{ij} - J) + p \delta(J_{ij}) + J) \right)+ q \delta(J_{ij}).
\end{equation}
Here, $p$ is the concentration of the antiferromagnetic interactions and $q$ denotes the ratio of the bonds.
We consider the case with bond dilution.
In the limit $p\to 0$, our model is reduced to the bond-diluted Ising model and exhibits the phase transition at the percolation threshold $q_c=1/2$ in the ground state.
In the limit $q\to 0$, our model becomes the $\pm J$ Ising model and shows the phase transition at the nontrivial point $p_c = 0.8967$ in the ground state.
The nature of the phase transition in the low-temperature region stems from the geometry of the frustration defined as $f_P = {\rm sign} \prod_{(ij) \in P}J_{ij}$ for each plaquette $P$.
The estimation of the critical points in the ground state proposed by Miyazaki computes a special quantity consisting of the expectation of the antiferromagnetic interactions on the lattice $N_a(p)$ and that of the frustration on the lattice $N_f(p)$ defined as
\begin{equation}
v(p) = \left( \frac{dN_a(p)}{dp}\right)^{-1}\frac{dN_f(p)}{dp}
\end{equation}
In the case of the square lattice, $v(p) = 2(1-2p)^3$.
The quantity $v(p)$ increases following the induction of the antiferromagnetic interactions.
Miyazaki conjectured that $v(p_c) = 1$ at the critical point in the ground state.
The resulting estimation $p_c=0.8969$ is very close to the existing result $p_c=0.8967(1)$ given by the numerical calculations \cite{Miyazaki2013}.

\begin{figure}[h]
\begin{center}
\includegraphics[width=0.48\textwidth]{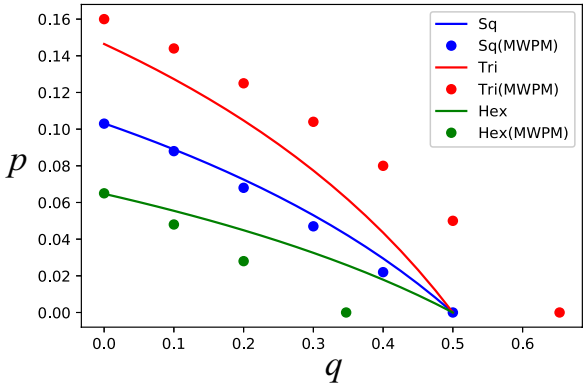}
\end{center}
\caption{(Color Online) Estimations (lines) and numerical results (circles) \cite{Stace2010} of the critical points in the ground state of the $\pm J$ Ising model with bond dilution on the square (blue and center), triangular (red and upper), and hexagonal lattices (green and lower).}
\label{fig1}
\end{figure}

We extend the Miyazaki conjecture to the case with bond dilution.
The direct manipulation of the replacement of the concentration of the antiferromagnetic interaction as $p \to (1-q)p$ does not work well.
In addition, if we take the change of the lattice into account, the computation of the number of antiferromagnetic interactions and frustrations can be highly nontrivial.
Thus, we consider some approximate treatment of bond dilution.
The bond dilution cannot affect the rate of decrease of the number of the antiferromagnetic interactions.
Thus $dN_a(p)/dp$ does not change.
However, the number of frustrations can decrease due to the damage to the lattice caused by bond dilution.
We assume that the effect simply comes with the first order of $1-q$ as $dN_f(p)/dp \to  (1-q)dN_f(p)/dp$.
Here we do not estimate a coefficient appearing in this change.
Instead, we check that our extension can recover the bond-percolation thresholds in the limit of $p \to 0$.
A simple extension of the conjecture is thus raised for satisfying this demand as
\begin{equation}
v(p) = (1-q)\left( \frac{dN_a(p)}{dp}\right)^{-1}\frac{dN_f(p)}{dp}
\end{equation}
The computation to obtain the quantities on the right-hand side is performed without bond dilution.
In this equality, when $p \to 0$, the bond-percolation threshold can be reproduced by setting $v(p_c) = 1$.
In addition, as in Fig. \ref{fig1}, several estimated values of the critical points in the ground state for various values of $q$ are shown.
We can confirm the reasonable coincidence between our estimations and the existing numerical results given by the Minimum-Weight Perfect Matching (MWPM) algorithm, which can compute the criticial points in the ground state of the $\pm J$ Ising model \cite{Stace2010}.
The extension is based on the consistency with the bond-percolation threshold on the square lattice, and thus fails to estimate the critical points on the triangular and hexagonal lattices.
Multiplication of an adequate coefficient $c$ as $dN_f(p)/dp \to  c(1-q)dN_f(p)/dp$ might lead to better expansion to the cases on the triangular and hexagonal lattices.
However we do not find any reasonable coefficient, which is consistent with the bond-percolation threshold in $p \to 0$ on these lattices, which should be clarified in future work.

\begin{figure}[h]
\begin{center}
\includegraphics[width=0.5\textwidth]{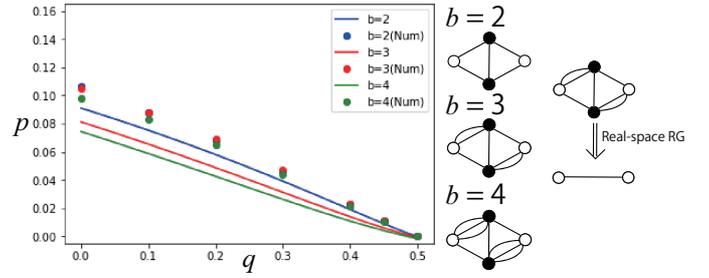}
\end{center}
\caption{(Color Online) Estimations (lines) and numerical results (circles) of the critical points in the ground state of the $\pm J$ Ising model with bond dilution on several self-dual hierarchical lattices.}
\label{fig2}
\end{figure}

We also test our extension in the cases of the self-dual hierarchical lattices, on which the bond-percolation threshold on them has the same value as that on the square lattice.
We compute the exact values of the critical point on several self-dual hierarchical lattices by performing the real-space renormalization group analysis \cite{Nobre2001}.
As shown in Fig. \ref{fig2}, the internal spins (black circles) are decimated to obtain a single bond with renormalized interaction.
The estimations of the critical points in the ground state by our extension slightly deviate from the numerical calculations as shown in Fig. \ref{fig2}.
Notice that the original estimation of the critical points in the ground state without bond dilution deviates the most from the numerical calculations \cite{Miyazaki2013}.
Although our equality became better as the degree of dilution $q$ increases, we do not assert the validity of our extension.
A severe problem in the application of the original theory based on the frustration might remain.

Summarizing these challenges, although we find the reasonable estimations of the critical points in the ground state of the $\pm J$ Ising model with bond dilution on the square lattice, our extension might show somehow accidental coincidence.
We however hope that our extension can provide clues for obtaining a reasonable theory to estimate the critical points even in the ground state of the finite-dimensional spin glasses.

\section*{Acknowledgement}
The present work was carried out with financial support from the JST-CREST(No.JPMJCR1402), and JSPS KAKENHI No. 15H03699 and Inamori Foundation.

\bibliography{paper_ver2}
\end{document}